\documentclass[12pt]{iopart}
\usepackage[utf8]{inputenc}
\usepackage{graphicx}
\usepackage{todonotes}
\usepackage{siunitx}
\usepackage{subcaption}
\usepackage{soul}
\usepackage{booktabs}
\date{December 2022}

\begin{document}
\renewcommand{\thefootnote}{\arabic{footnote}}

\title[RFKO Extraction Pulses]{Pulsed RF Knock-Out Extraction: A Potential Enabler for FLASH Hadrontherapy in the Bragg Peak}
\author{Simon Waid$^1$, Andreas Gsponer$^{1,2}$, Elisabeth Renner$^2$, Claus Schmitzer$^3$,  Florian Kühteubl$^2$,   Clara Becker$^2$, Jürgen Burin$^1$, Philipp Gaggl$^1$, Dale Prokopovich$^3$, Thomas Bergauer$^1$}

\address{$^1$ Austrian Academy of Sciences, Institute for High Energy Physics, Vienna, Austria}
\address{$^2$ Vienna University of Technology, Vienna, Austria}
\address{$^3$ MedAustron GmbH, Wiener Neustadt, Austria}

\vspace{10pt}
\begin{indented}
\item[]November 2023
\end{indented}

\begin{abstract}
   One challenge on the path to delivering FLASH-compatible beams with a synchrotron is facilitating an accurate dose-control for the required ultra-high dose rates. We propose the use of pulsed RFKO extraction instead of continuous beam delivery as a way to control the dose delivered per Voxel. In a first feasibility test dose rates in pulses of up to \qty{600}{\gray\per\second} were observed, while the granularity at which the dose was delivered is expected to be well below \qty{0.5}{\gray}.
\end{abstract}

\section{Introduction}

The FLASH effect was first described in the 1960s and 1970s \cite{berryEFFECTSRADIATIONDOSERATE1973, hornseyUnexpectedDoserateEffect1966}. However, it did not find clinical application up until recently when it was re-discovered and put into practical use during the last decade 
\cite{hughesFLASHRadiotherapyCurrent2020a, favaudonUltrahighDoserateFLASH2014}. 
Since then, FLASH radiotherapy has seen increased research interest. Initial research focused on electron and photon radiotherapy. Later, the existence of the FLASH effect was also confirmed for proton radiotherapy \cite{ashrafDosimetryFLASHRadiotherapy2020, velalopoulouFLASHProtonRadiotherapy2021}. Currently, a first study on humans using proton radiotherapy is ongoing \cite{tedeschiFLASHRadiationTherapy2022, chowFLASHRadiationTherapy2021}. The use of other ion species such as carbon is currently under investigation \cite{weberFLASHRadiotherapyCarbon2022b}.

Attaining FLASH conditions while using the Bragg peak for treatment has proven difficult. As a consequence, researchers often resort to irradiation in transmission mode, positioning the Bragg peak outside of the target. The transmission mode has been applied to the vast majority of in-vivo and in-vitro studies on FLASH radiotherapy using protons so far \cite{ashrafDosimetryFLASHRadiotherapy2020, velalopoulouFLASHProtonRadiotherapy2021, tedeschiFLASHRadiationTherapy2022, chowFLASHRadiationTherapy2021, weiFLASHRadiotherapyUsing2022, diffenderferCurrentStatusPreclinical2022}. 

Positioning the Bragg peak outside the tumor volume mitigates one of the main advantages of proton radiotherapy: The larger dose deposition  
in the Bragg peak compared to the entrance channel \cite{mohanReviewProtonTherapy2022a}. When moving FLASH radiotherapy from research to the clinic, weighting the benefits of the FLASH-effect against the benefits of using the Bragg peak will become a difficult endeavour. Combining the two effects would be highly desirable.

The difficulties encountered when targeting FLASH conditions while using the Bragg peak for treatment are mainly related to the required modification of the beam energy to position the Bragg peak in the tumor volume: (i)  dilated treatment times due to switching times between energy layers and (ii) the limited beam flux at low beam energies in cyclotrons. Both issues have been addressed for cyclotrons using a combination of universal range-shifters and field-specific range compensators \cite{kangUniversalRangeShifter2022, weiNovelProtonPencil2021, weiAdvancedPencilBeam2022}. Thus, for cyclotrons, FLASH beams using the Bragg peak can be deemed feasible even if further development might be needed prior to clinical application. For synchrotron accelerators, irradiation using the Bragg peak at FLASH dose rates has been demonstrated for shallow tumors in mice \cite{dokicNeuroprotectiveEffectsUltraHigh2022}. Similar to cyclotrons, the use of range compensators can be expected to enable a spread out Bragg peak.

One issue still existing with synchrotron accelerators, both when targeting treatment in transmission mode or using the Bragg peak, is beam monitoring and beam control due to the highly fluctuating beam intensity. In particular, for synchrotron accelerators it is challenging to keep the dose delivered under fault conditions within acceptable limits.  



Safety concepts for synchrotron accelerators typically rely on monitoring the beam parameters and interrupting or terminating irradiation if the parameters are out of specifications. Detection of deviations and termination has to happen before an intolerable dose is delivered to the patient. The relevant tolerable dose for each detection and termination is \qty{0.25}{\gray} (IEC 60601-2-64, 2014, clause 201.10.2.101.3.1.6). When driving extraction at the maximum available extraction speed, one can assume the dose is proportional to the extraction time. 
If the dose delivered within a single  short, FLASH-compatible high-dose pulse is below this limit, it becomes acceptable to have no means of termination during this pulse. Still, after such a pulse and prior to any subsequent pulse, an evaluation of the delivered dose and beam parameters needs to be carried out. This verification may then lead to a termination of irradiation.

\begin{figure}[htp]
    \centering
    \includegraphics{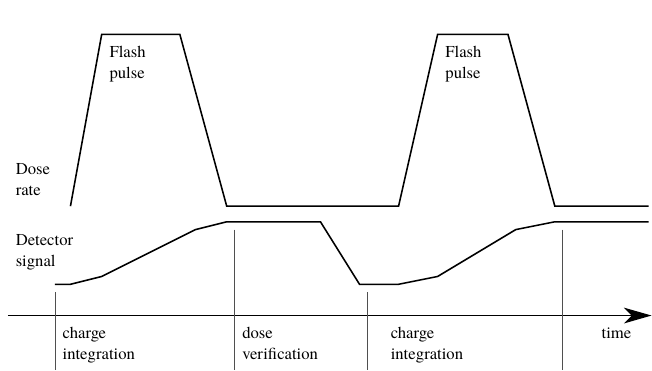}
    \caption{Synchronization between extraction and beam monitors. While the beam is being extracted with FLASH-compatible dose rates, the beam monitors integrates the charge from the detector. During the verification of the beam parameters, no beam is extracted. If the beam parameters are out of limits, the subsequent extraction pulse is inhibited and irradiation is terminated.}
    \label{fig:synchronization}
\end{figure}

For optimum usage of beam monitors, extraction and beam monitor readout can be synchronized. An approach for  synchronization is given in fig. \ref{fig:synchronization}. The extraction is performed in short pulses instead of continuously. The beam monitoring is carried out in two phases: A measurement and a verification phase. The measurement phase needs to be carried out during the delivery of a pulse. The verification can be carried out after the pulse has been delivered.  E.g. the beam monitors may integrate the readings from a detector during a pulse in a fast, analog memory. In a short pause between pulses, the values can be digitized and verified. Delivery of the next pulse is only permitted in case the verification shows the pulse was within limits. Depending of the requirements for FLASH-compatible beams a pulse-train can be administered with the aim of providing an average dose rate as well as accumulated dose above the FLASH limit. 


A promising technique to enable extraction short pulses is radio frequency knock out (RFKO) extraction, which is already applied in several ion therapy synchrotrons for nominal operation \cite{hiramotoResonantBeamExtraction1992, savazziImplementationRFKOExtraction2019, krantzSlowExtractionTechniques2018, feldmeierUpgradeSlowExtraction2022} and explored as alternative extraction mechanism at MedAustron~\cite{floriankuhteublDesignStudyRadio2020}.
This slow extraction method applies a horizontal electric RF field to increase the amplitude of the horizontal particle oscillation around the reference orbit until they reach the unstable region around the resonance and get extracted. 

In this work we propose a beam delivery concept utilizing short extraction pulses. We believe this short pulses can aid in complying with safety regulations, as they can aid in optimally profiting from the bandwidth of the beam monitors.


\section{Material and Methods}
The presented measurements were performed at the synchrotron of the MedAustron facility located in Wiener Neustadt, Austria. The accelerator is based on the
proton-ion medical machine study (PIMMS) design \cite{bryantProgressProtonIonMedical1999} and enables proton energies from \qty{62.4}{\mega\electronvolt} to \qty{800}{\mega\electronvolt}, whereby energies up to \qty{252.7}{\mega\electronvolt} are employed clinically. When configured for clinical operation, the accelerator extracts the beam using a Betatron core. However, for experiments, the accelerator can be configured to extract the beam using alternative extraction methods such as constant optics slow extraction (COSE) \cite{kainResonantSlowExtraction2019, arrutiasotaImplementationTuneSweep2022} or RFKO \cite{vandermeerStochasticExtractionLowripple1978}. The RFKO set-up at MedAustron  \cite{wastlalexanderInvestigatingAlternativeExtraction23} was still under development during the presented proof-of-concept pulsed RFKO extraction tests.  The applied  machine settings  were hence not optimized for RFKO extraction, but rather for an extraction based on momentum selection (betatron core extraction). The estimated settings are listed in tab.~\ref{tab:machinesettings}. For our experiment we chose an unbunched (coasting) proton beam with a constant beam energy of \qty{252.7}{\mega\electronvolt} and a ring filling of $0.9 \cdot 10^{10}$  to $1.3 \cdot 10^{10}$ protons after acceleration.

The RFKO signal was generated using an Ettus USRP X310 software defined radio (SDR) and amplified using a custom built \qty{1}{\kilo\watt} amplifier. The transverse excitation of the beam was achieved by feeding the amplified signal via a custom built BalUn to the plates of the Schottky beam monitor~\cite{wastlalexanderInvestigatingAlternativeExtraction23}. 
The SDR was configured to emit periodic pulses at a rate of \qty{100}{\hertz} with 50\,\% duty cycle. 

The excitation signal was set to a base frequency of \qty{3.964}{\mega\hertz}, which is chosen to correspond to the expected horizontal betatron tune of $Q_x \approx 1.679\pm 4 \times 10^{-4}$ (95\% confidence level).
The  horizontal chromatic betatron tune spread is estimated to be Gaussian with a FWHM of $\Delta Q_{x,\text{FWHM}}\approx 0.0032$, corresponding to a frequency spread of $\Delta f_\text{FWHM} \approx 7.5$\,kHz. The RFKO excitation frequency was modulated by a sawtooth signal with a sweep time of 10\,ms and a frequency modulation amplitude of $\Delta f \approx \SI{800}{\hertz}$. As the frequency modulation could not be synchronized to the RFKO pulse gating, it is not possible to state the exact resulting base frequency and bandwidth but rather confine the effective bandwidth to be within the range of $\Delta f \approx 400\text{-}\SI{800}{\hertz}$. The resultant uncertainty can be neglected for this study, as it is an order of magnitude smaller than the frequency spread of the beam. 
The excitation amplitude was varied between extractions by changing the output level on the SDR between 0.04375 and 0.7 relative to the maximum permitted by the SDR. 

\begin{table}[]
\centering
\begin{tabular}{@{}llc@{}}
\toprule
\textbf{Parameter}         & \textbf{Unit}                                                                & \textbf{Value}     \\ \midrule
Nominal $Q_{x,0}$  & -                                      & 1.669  \\
Horizontal chromaticity $Q'_x$   & -                                      & -4.1                 \\
Relative momentum offset and resulting $Q_x$ & - / -                                       & -0.0025 and 1.679  \\
RF settings, longitudinal distribution    & -                                                & Coasting (unbunched) \\
Resonant sextupole strength k$_2$L &  m$^{-2}$                           & 2.2            \\
   &                                                  &                  \\
Intensity before extraction    & protons                                                           & $0.9$-$1.3\times$\SI{e10}{}      \\
Relative momentum spread (FWHM)    & -                                                           & \SI{8e-4}{}         \\
Normalized horizontal rms emittance $\epsilon_{\text{n, rms},x}$ & mm mrad & 0.5                     \\
Normalized vertical rms emittance $\epsilon_{\text{n, rms},y}$ & mm mrad   & 0.5                  \\ 
Revolution time of synchronous particle &  \si{\micro\second}                           & 0.423            \\ \bottomrule
\end{tabular}
\caption{Estimated machine and beam parameters during the presented pulsed RFKO tests (flat-top, prior to extraction). }\label{tab:machinesettings}
\end{table}

\begin{figure}[htp]
    \centering
    \includegraphics[width=0.8\textwidth]{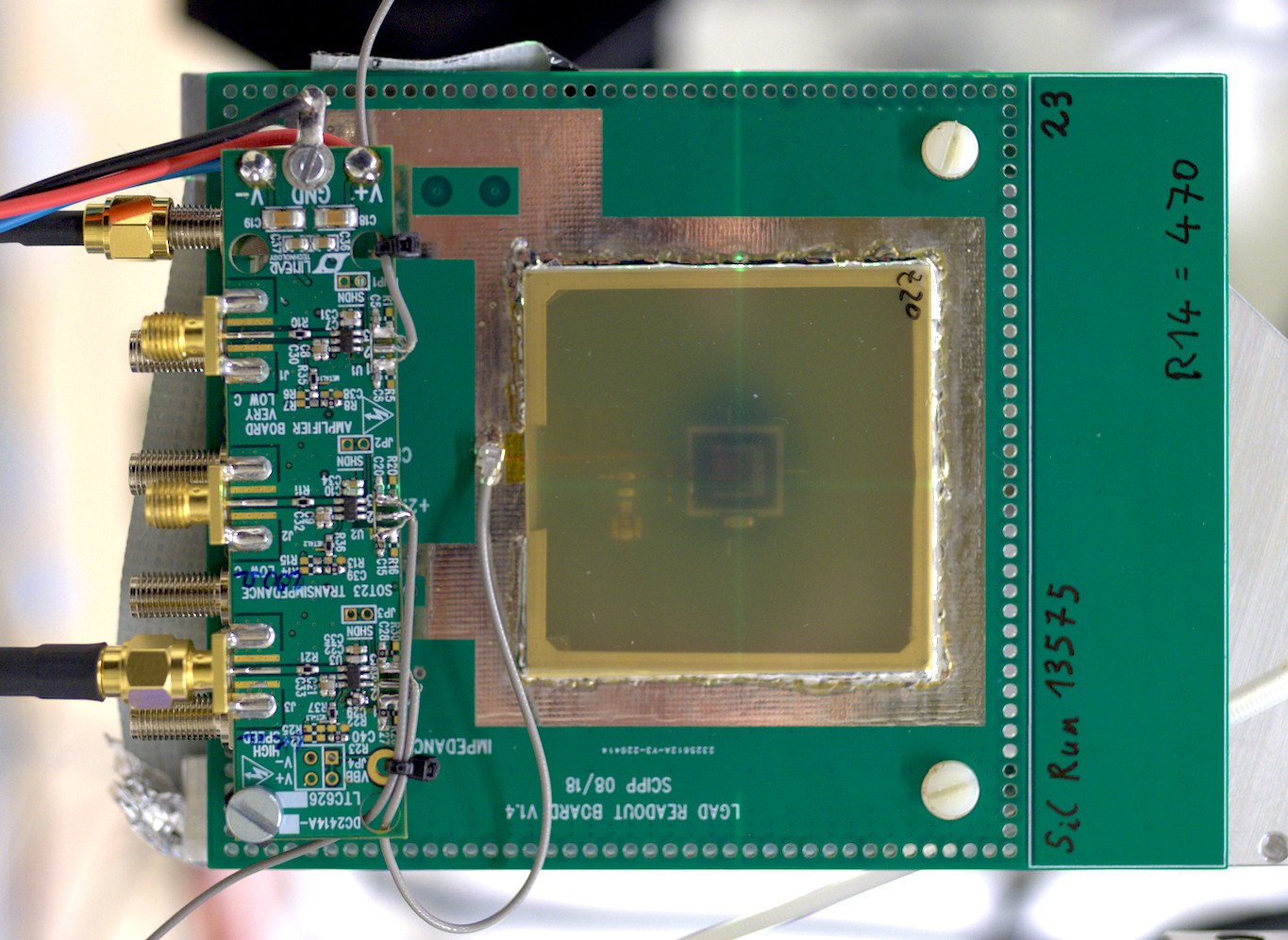}
    \caption{Measurement setup with mounted EBT3 film as it was employed to measure the absolute dose. The film was glued to a metal frame surrounding the SiC detector using double-sided tape. 
    }
    \label{fig:meas-setup}
\end{figure}

Dosimetry was carried out using a combination of the current transformer in the synchrotron and EBT3 films  and 4H-SiC detectors in the irradiation room. EBT3 films and the current transformers were used for absolute dosimetry by measuring the dose and intensity, respectively. The 4H-SiC detector measured the time structure of the dose rate. At the beginning of the experiment a reference measurement was performed using one EBT3 film. This measurement was employed to link the change in current in the ring during extraction and the integrated detector current to a delivered dose. The EBT3 film was scanned prior to exposure and positioned in the iso-center for exposure as shown in fig. \ref{fig:meas-setup}. Post-exposure, the film was aged for \qty{48}{\hour} and scanned again. A Co-60 derived calibration curve was employed to convert the darkening of the film to an integral dose. The time structure of the dose rate during each extraction was computed by normalizing the current from the 4H-SiC detector to the dose determined via the corresponding current measurement in the ring.

4H-SiC detectors were provided by Centro Nacional de Microelectrònica (CNM), Barcelona, Spain. The detectors were fabricated on a 4H-SiC substrate with a \qty{50}{\micro\meter} thick epitaxially grown, n-doped active layer and had an active area of 3x3 \unit{\square\milli\meter}. More details on the employed detectors are given in \cite{christanell4HsiliconCarbideParticle2022, gagglChargeCollectionEfficiency2022, gagglPerformanceNeutronirradiated4HSilicon2022a}. The detector was positioned in iso-center. Centering of beam was verified via EBT3 film measurements in order to minimize the impact of lateral beam movements on the measurement. The beam had a full width at half maximum FWHM of \qty{9.5}{\milli\meter}.  A Keithley 2470 SMU was employed to bias the detector and measure the current through the detector. The sampling rate of the SMU was configured to the maximum available frequency of \qty{2.8}{\kilo\hertz}. 

In addition to the SMU, the current through the detector was also measured using a custom made transimpedance amplifier (TIA). The TIA consisted of an Analog Devices LTC6268-10 operational amplifier with a single feedback resistor as well as a compensation capacitor. The TIA had a transmittance of \qty{46}{\decibel\ohm} 
and a bandwidth of \qty{20}{\mega\hertz}. The output of the TIA was digitized using a Rohde \& Schwarz RTP164 oscilloscope configured in HD mode to attain an ADC resolution of \qty{16}{\bit}. The sampling rate of the oscilloscope was between \qty{100}{\mega\hertz} and \qty{200}{\mega\hertz}. The entire \qty{10}{\second} chopper opening time were recorded and stored in the oscilloscope memory.

\section{Results and Discussion}

\begin{figure}[htp]
    \centering
    \includegraphics[width=0.9\textwidth]{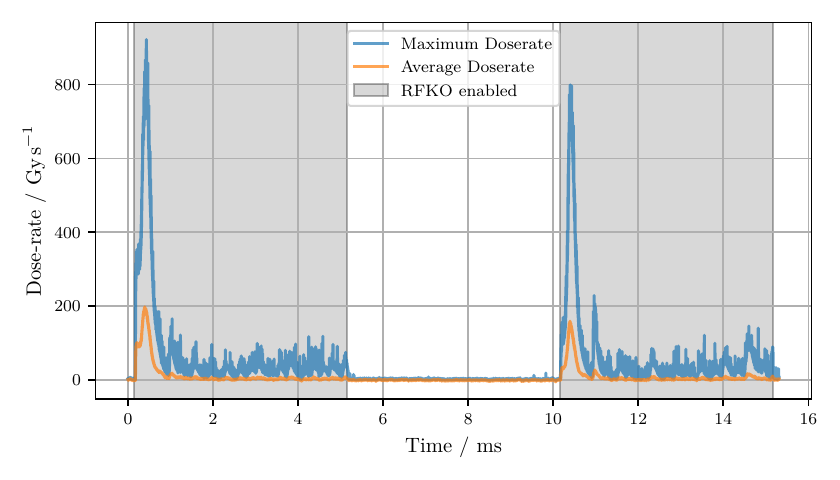}
    \caption{Dose rate over time for a RFKO extraction at a gain factor of 0.175. The gray area between vertical black lines denote the time when the RFKO signal was enabled. The average and maximum were calculated over a sliding window of 100 samples. The moving average shows a distinct initial extraction peak at high dose rate, followed by a 20 to 30 times lower dose rate.}
    \label{fig:ko-dr-time}
\end{figure}

The measurement result for a \qty{5}{\milli\second} pulse with a SDR gain factor of 0.175 is displayed in fig. \ref{fig:ko-dr-time}. The areas coloured in gray indicate the time intervals during which the beam gets excited by the RFKO signal. Since the detector bandwidth exceeds the frequency of the RFKO excitation, dose rates were fluctuating substantially. For better clarity, the plot shows the average (blue) and the peak dose rate (orange) calculated as the moving average dose rate over 100 samples, corresponding to averaging time of \qty{1}{\micro\second}. 

The measured dose rate profile is shown in fig. \ref{fig:ko-dr-time}. After the RFKO signal is enabled, the dose rate rises to a maximum, which is reached after \qty{260}{\micro\second}. However, after reaching the maximum, the dose rate falls rapidly and reaches a steady state which is 20 to 30 times lower than the peak dose rate even though the RFKO extraction signal is still being applied.

\begin{figure}[htp]
     \centering
     \includegraphics[width=0.9\textwidth]{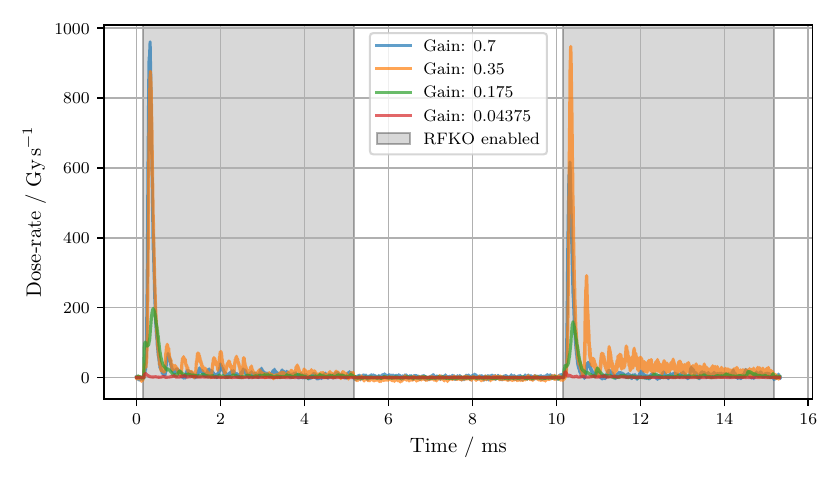}
     \caption{Average dose rate over time for different gains. The duration of the pulse stays almost constant, while the height of the pulse scales with the applied gain. 
     }
     \label{fig:doserate}
 \end{figure}

Dose rates over time for different gain settings are shown in fig. \ref{fig:doserate}. Peaks in the average dose rates of up to \qty{900}{\gray\per\second} can be observed, while the dose rate dropped substantially after an initial peak. Assuming a pulse duration of \qty{0.5}{\milli\second} this corresponds to a dose rate of \qty{600}{\gray\per\second} at our detector.  

\begin{figure}[htp]
    \centering
    \includegraphics[width=0.8\textwidth]{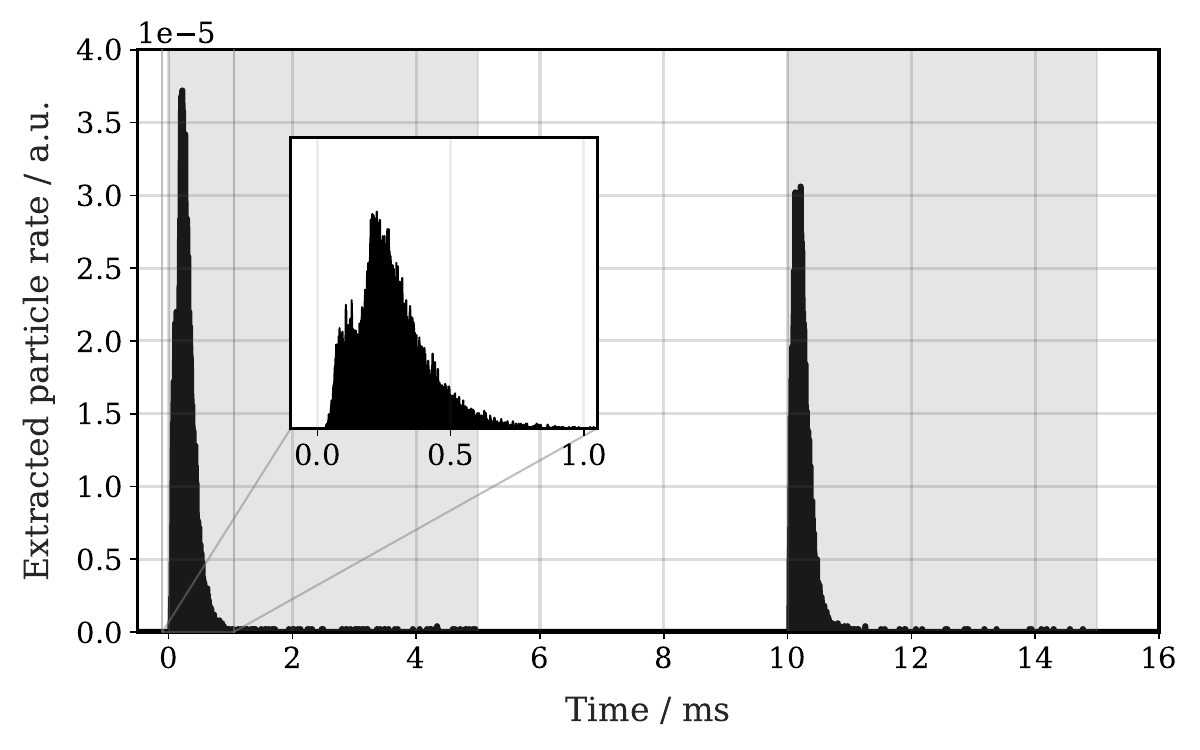}
    \caption{Simulated rate of extracted particles $dN_\text{extr.}/dt$. The grey shaded areas denote the time intervals of the RFKO excitation.}
    \label{fig:mad-x}
\end{figure}

We reproduced the measurements using Xsuite \cite{iadarolaXsuiteIntegratedBeam2023}  simulations. 
While these simulations serve as a first qualitative comparison, it is important to acknowledge that further comprehensive analysis,  sensitivity studies and beam measurements are required to  enable a more detailed and also quantitative comparison between simulation and measurements in future studies. 

The simulated rate of extracted particles $dN_\text{extr.}/dt$ is illustrated in fig.~\ref{fig:mad-x}. The characteristic of the initial peak per pulse, in which the particles close to the separatrix are extracted, aligns well in simulations and measurements, both featuring a width of $\approx 0.5\,$ms. The beam dynamics in play during the presented pulsed narrow-bandwidth RFKO excitation is currently under investigation. We refer the reader to the work of P. Niedermayer and R. Singh in \cite{niedermayerExcitationSignalOptimization2024}, which provides an excellent analysis on the motion of particles under the influence of a sinusoidal excitation, as well as the work of M. Pullia in \cite{pulliaTimeProfileSlowly1997}, which provides a detailed analytical analysis of the time profile of a series of particles with different momenta and amplitude, that become unstable simultaneously.

\begin{figure}[htp]
    \centering
    \includegraphics[width=0.9\textwidth]{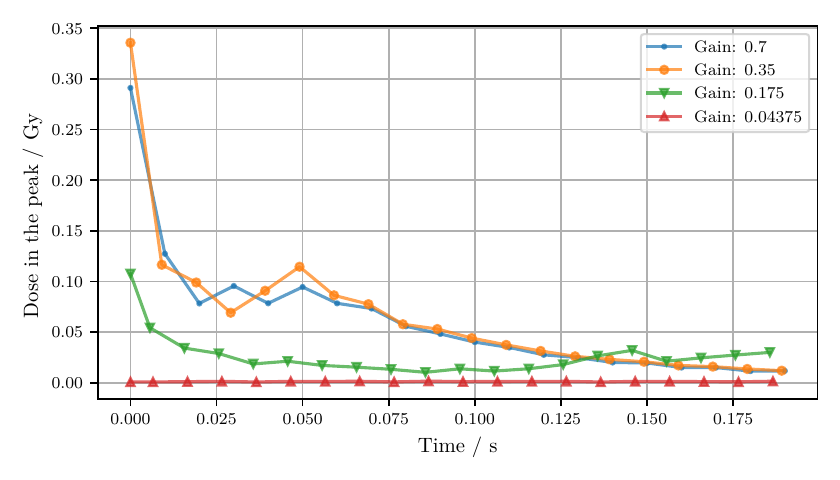}
    \caption{Measured dose in each peak over time. 
    }
    \label{fig:ko-dose-peak}
\end{figure}

We further analyzed the measured dose in the initial dose rate spike for each peak. The resulting dose per peak is shown in fig. \ref{fig:ko-dose-peak}. The first extraction pulse shows a significantly larger extracted dose than the subsequent pulses. After this first pulse, the dose per pulse drops significantly and is almost constant before it falls after approximately 15 peaks. 
Optimizing  the machine settings and excitation signal as well as modulating the gain pulse-to-pulse is expected to help harmonizing the delivered dose rate between pulses. Fig. \ref{fig:ko-dose-peak} also shows that for a gain of 0.35 and 0.7 the dose extracted per peak is almost independent from the applied gain factor. This might be due to reaching the power limit of the employed RF-amplifier. For lower gains this is not the case.


The observed effect can be beneficial in ensuring patient safety. In contrast to only limiting the extracted dose via the RFKO on-time,  the observed peak-tail characteristic gives time to safety system to react to a malfunction. 
If the RFKO signal is applied for too long, the dose rate will still drop. Safety systems now have time to detect the failure, react and terminate irradiation.

\begin{figure}[htp]
     \centering
     \includegraphics[height=0.3\textheight]{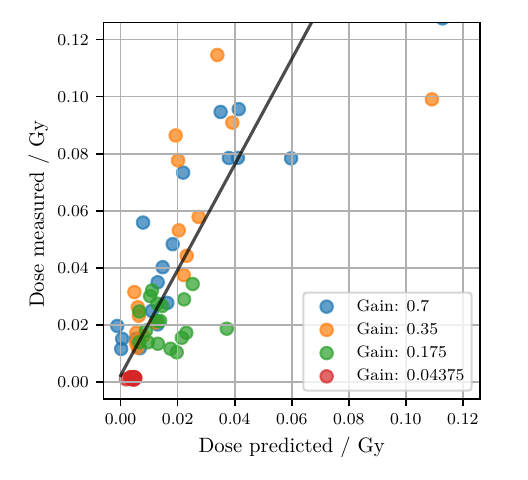}
     \caption{Relationship between the dose in the initial peak of a pulse and the dose expected when integrating the average dose rate in the flat part of the previous pulse over the pause between RFKO extraction pulses. The line indicates a linear relationship and helps to guide the eye. The measured dose in the peak is approximately twice the with said ansatz computed dose.
     }
     \label{fig:ko_peak_replenishment}
 \end{figure}

We tried to predict the extracted dose in the peak of each pulse from the dose rate measured in the tail of the previous pulse, 
 making the ansatz that the dose in the peak of the  $(n+1)$-st pulse corresponds approximately to the flat dose rate in the $n$-th pulse multiplied by the time between pulses. 
 The results are illustrated in Fig. \ref{fig:ko_peak_replenishment},
 indicating a linear correlation. Interestingly, the measured dose is almost twice as large as the predicted dose. 

\begin{figure}[htp]
    \centering
    \includegraphics[width=0.9\textwidth]{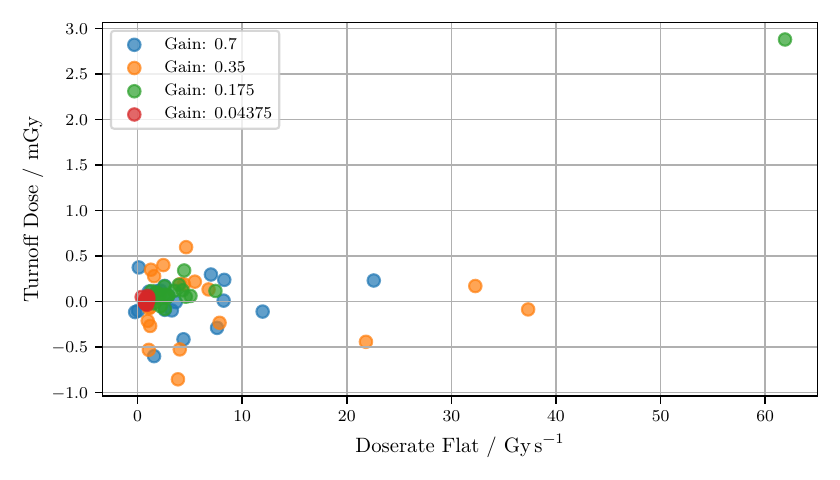}
    \caption{Computed dose delivered during termination. The plot shows the dose delivered within 170~µs after the RFKO signal was switched off as a function of the dose rate during the flat part of the extraction. The data is dominated by noise. The dose is below \qty{1.1}{\milli\gray} in all cases. 
    }
    \label{fig:ko_turnoff}
\end{figure}

In case of the MedAustron accelerator  termination of a treatment at the MedAustron facility is primarily attained via the chopper system \cite{pulliaBetatronCoreDriven2016}. For protons, the system has a maximum response time of \qty{150}{\micro\second}. In addition to the chopper closing time, we assess there will be the need to process the measurement data from the extraction pulse. We assess, this processing will take less than \qty{20}{\micro\second}. Thus, to calculate the expected termination dose, we can calculate the dose delivered during the \qty{170}{\micro\second} after switching off the RFKO signal. The result of this calculation is given in fig. \ref{fig:ko_turnoff} as a function of the average dose rate during the flat part of the extraction pulse. Due to the low dose rates during this time-frame, the data is noisy and the calculation yields even nonphysical negative doses. The data shows that the dose is below \qty{3}{\milli\gray} in all cases. 

\section{Conclusion and Outlook}
We have shown that pulsed RFKO extraction can be used to attain extraction pulses with large instantaneous dose rates but a limited dose per pulse due to the short duration of the pulses. This can be utilized to optimize the available bandwidth of beam monitors by synchronizing the beam monitors to the RFKO excitation. In our experiments, we recorded dose rates of up to \qty{600}{\gray\per\second}, assuming a pulse duration of \qty{0.5}{\milli\second}. This dose rates are still for single pulses. Further research on range modulation, scanning strategies and the extraction mechanism is necessary before the practically attainable dose per extraction and the practical dose averaged dose rate (DADR) can be determined.   

We observed that after turning on the RFKO signal for extremely large extraction rates, the measured dose rate drops by a factor of 20 to 30 after an initial peak. Further investigation of the beam dynamics at play is necessary to obtain a complete picture of this behavior. 
 The drop in extracted dose rate when operating the accelerator with pulsed RFKO extraction can be beneficial for ensuring safety. In case of a failure  the low dose rate in the pules tails gives safety systems time to react. 
 We believe that the presented extraction and beam monitoring approach can be attractive for the delivery of scanned FLASH beams using synchrotron accelerators. 

\section{Acknowledgement}
This project has received funding from the Austrian Research Promotion Agency FFG, grant number 883652. Production and development of the 4H-SiC detector was supported by the Spanish State Research Agency (AEI) and the European Regional Development Fund (ERDF), ref. RTC-2017-6369-3.
\section{Bibliography}

\bibliography{HEPHY-detector-dev} 
\bibliographystyle{ieeetr}

\end{document}